\documentstyle[12pt,aaspp4]{article}

\def\s{{\rm\,s}}

\def\cm{{\rm\,cm}}
\def\mm{{\rm\,mm}}
\def\gm{{\rm\,g}}

\def\au{{\rm AU}}
\def\mum{\,\mu{\rm m}}
\def\K{{\rm\,K}}
\def\yr{{\rm\,yr}}
\def\Hz{{\rm\,Hz}}
\def\AU{{\rm\, AU}}

\begin{document}

\lefthead{Chiang and Goldreich}
\righthead{T TAURI STARS WITH PASSIVE DISKS}

\title{SPECTRAL ENERGY DISTRIBUTIONS OF T TAURI STARS WITH PASSIVE
CIRCUMSTELLAR DISKS}

\author{E.~I.~Chiang and P.~Goldreich}

\affil{California Institute of Technology\\
Pasadena, CA~91125, USA}

\authoremail{echiang@tapir.caltech.edu and pmg@nicholas.caltech.edu}

\begin{abstract}
We derive hydrostatic, radiative equilibrium models for passive disks
surrounding T Tauri stars. Each disk is encased by an optically thin
layer of superheated dust grains. This layer re-emits directly to space
about half the stellar energy it absorbs. The other half is emitted
inward and regulates the interior temperature of the disk. The heated
disk flares. As a consequence, it absorbs more stellar radiation, especially
at large radii, than a flat disk would. The portion of the SED contributed
by the disk is fairly flat throughout the thermal infrared. At fixed frequency,
the contribution from the surface layer exceeds that from the interior by about
a factor 3 and is emitted at more than an order of magnitude greater
radius. Spectral features from dust grains in the superheated layer appear in
emission if the disk is viewed nearly face-on.
\end{abstract}

\keywords{circumstellar matter --- radiative transfer --- stars: pre-main
sequence --- infrared: stars --- accretion, accretion disks --- stars:
individual (GM Aur, GG Tau)}

\section{INTRODUCTION}

Excess infrared (IR) emission from T Tauri stars is thought to arise
from circumstellar disks \markcite{m68,sal87,als87}(Mendoza 1968; Shu, Adams \&
Lizano 1987;
Adams, Lada, \& Shu 1987; and references therein). The standard disk is
modeled as a flat blackbody. In the simplest scenario it lacks
intrinsic luminosity and passively reradiates the energy it absorbs
from the central star.  This blackbody model yields an infrared spectral energy
distribution (SED) of the form $\nu F_{\nu } \propto \nu ^n$, with
spectral index $n=4/3$ \markcite{als87} (Adams, Lada, \& Shu 1987; and
references
therein). By coincidence, the steady-state accretional luminosity from
an opaque, thin disk yields an SED with identical spectral index
\markcite{lp74} (Lynden-Bell \& Pringle 1974). SEDs measured between $3$ and
$100\mum$
are usually well fitted by power laws with $0\leq n\leq 4/3$. However, in
contrast to the prediction of the standard model, most sources exhibit
flattish spectra having $n\leq 3/4$ \markcite{retal84,r85,rz87,setal89,betal90}
(see, e.g., Rydgren et al. 1984;
Rucinski 1985; Rydgren \& Zak 1987; Strom et al. 1989; Beckwith et al.
1990).

The failure of the standard disk model spawned a number of alternative
proposals. Of greatest relevance here is the investigation by
\markcite{kh87}Kenyon \&
Hartmann (1987) of a blackbody disk whose surface flares outward with
increasing radius as a consequence of vertical hydrostatic equilibrium.
Flared disks intercept more stellar radiation than flat ones, especially at
large distances from the star. Other models invoke either an ``active''
disk having a high intrinsic luminosity, or a dusty component in
addition to the disk. The dissipation of a one-armed spiral density wave
as an intrinsic disk heating mechanism falls in the first category
\markcite{ars89,setal90,osa92}(Adams, Ruden, \& Shu 1989;
Shu et al. 1990; Ostriker, Shu, \& Adams
1992). The second category includes the proposal that a tenuous dusty
envelope surrounds the star/disk system and either scatters stellar
radiation back onto the disk \markcite{n93} (Natta 1993), or absorbs and
re-emits it in the
infrared \markcite{cetal94} (Calvet et al. 1994).

We calculate the SED for the passive, reprocessing disk in a
self-consistent fashion. The basic features and results of
our investigation are set forth in \S 2. Extensions and refinements are
considered in \S 3. Lastly, in \S4, we discuss our results in the
context of observations and summarize unresolved issues.

\section{THEORY}

\subsection{Model Assumptions}
\label{assume}

We consider a T Tauri star surrounded by a passive disk. The star
is modeled as a spherical black body of temperature $T_*= 4000\K$,
mass $M_*=0.5 M_\odot$, and radius $R_*= 2.5 R_{\odot}$
(see, e.g., Table II in \markcite{betal90} Beckwith et al. 1990).
Our fiducial disk has surface mass density similar to that of
the minimum-mass solar nebula, $\Sigma = a_\au^{-3/2}\Sigma _0$,
where $a_\au$ is the disk radius measured in AU and $\Sigma _0 =
10^3\gm\cm^{-2}$ \markcite{w77} (Weidenschilling 1977). Dust, which is
uniformly
mixed with the gas, comprises about one percent of the total mass. Dust grains
dominate the continuum opacity from visible through millimeter wavelengths.
We take the grains to be spheres with radius $r = 0.1\mu$m, mass density
$\rho_d = 2\gm\cm^{-3}$, and negligible albedo. Their emissivity, which is
unity for $\lambda\leq 2\pi r$, decreases
as $\varepsilon_\lambda\approx (2\pi r/\lambda)^{\beta }$ for $\lambda\geq
2\pi r$. We denote by $\varepsilon\approx (8\pi rkT/hc)^{\beta }
\approx(T/T_*)^{\beta }$
the average dust emissivity at temperature $T$. For our fiducial disk,
$\beta = 1$. The dust opacity at visual wavelengths is
$\kappa _V \approx 400 \cm^2 \gm^{-1}$, which implies an optical depth
$\tau_V\approx 4 \times 10^5 \; a_\au^{-3/2}$ for our fiducial disk.

SEDs are computed for disks viewed pole-on.  We choose an inner cutoff
radius, $a_i\approx 6R_*\approx 0.07\au$, to mark the condensation
boundary of common silicates. The outer cutoff radius for flat disks,
$a_o\approx 2.3\times 10^4R_*\approx 2.7\times 10^2\au$, is fixed to facilitate
comparisons among different disk models; $a_o$ is comparable to the
size of the largest disks seen in silhouette against the Orion nebula
\markcite{mo96} (McCaughrean \& O'Dell 1996).

Our model assumptions are chosen to make the telling of our story as
simple and direct as possible. Issues regarding these choices and
possible alternatives are dealt with in \S 3.

\subsection{The Blackbody Disk}

To begin, we review standard relations for the temperature and SED of a
blackbody disk.  The flux of stellar radiation incident upon the disk is
$(\alpha/2)\; (R_*/a)^2\sigma \; T_{\ast}^4$ for $a\gg R_*$, where $\alpha$ is
the grazing angle at which the starlight strikes the disk.
Equating the emitted and absorbed fluxes yields the
disk temperature\footnote{The subscript $e$ is used to denote effective
temperature.}

\begin{equation}
T_e \approx  \left(\frac{\alpha}{2}\right)^{1/4}\;
\left(\frac{R_*}{a}\right)^{1/2}\; T_* \; .  \label{Te}
\end{equation}

The SED is computed as

\begin{equation}
L_{\nu} \equiv 4\pi d^2\nu F_{\nu} = 8\pi^2\nu \int_{a_i}^{a_o}\, da\, a
B_{\nu}(T_e)\, , \label{bbsed}
\end{equation}

\noindent where $B_{\nu}(T)$ is the Planck
function, and $d$ is the distance to the source. A scaling relation for
$L_{\nu}$ at wavelengths between those that characterize the disk at $a_i$ and
$a_o$ is derived as follows. Equation (\ref{bbsed}) is
approximated as $8\pi^2a^2\nu B_{\nu}(T_e)$, $\nu B_{\nu}(T_e)$ is replaced by
$\sigma T_e^4 / \pi$, $a$ is related to $T_e$ by equation (\ref{Te}), and $T_e$
is expressed in terms of $\nu$ by $3kT_e\sim h\nu$. These steps yield

\begin{equation}
L_{\nu} \sim 8 \pi a^2 \sigma T_e^4 \sim \alpha L_* \, . \label{sedapprox}
\end{equation}

\noindent In other words, the fraction of the stellar luminosity $L_*$ that is
reprocessed to frequencies in an octave centered on $\nu$ is approximately
equal to the grazing angle $\alpha$ at the location where $3kT_e=h\nu$.

\subsubsection{Flat Geometry}
\label{bbflat}

A flat disk is one whose aspect ratio (opening angle) is independent of $a$; we
assume it to be much less than unity. The grazing angle appropriate to this
geometry is $\alpha\approx  0.4 R_*/a\ll 1$ for $a/R_*\gg 1$ (Kusaka, Nakano,
\& Hayashi 1970, Ruden \& Pollack 1991). Thus $T_e$ from equation (\ref{Te})
takes the form

\begin{equation}
T_e \approx  \left(\frac{2}{3\pi}\right)^{1/4}\;
\left(\frac{R_*}{a}\right)^{3/4} \; T_* \; . \label{Tef}
\end{equation}

\begin{figure}
\plotone{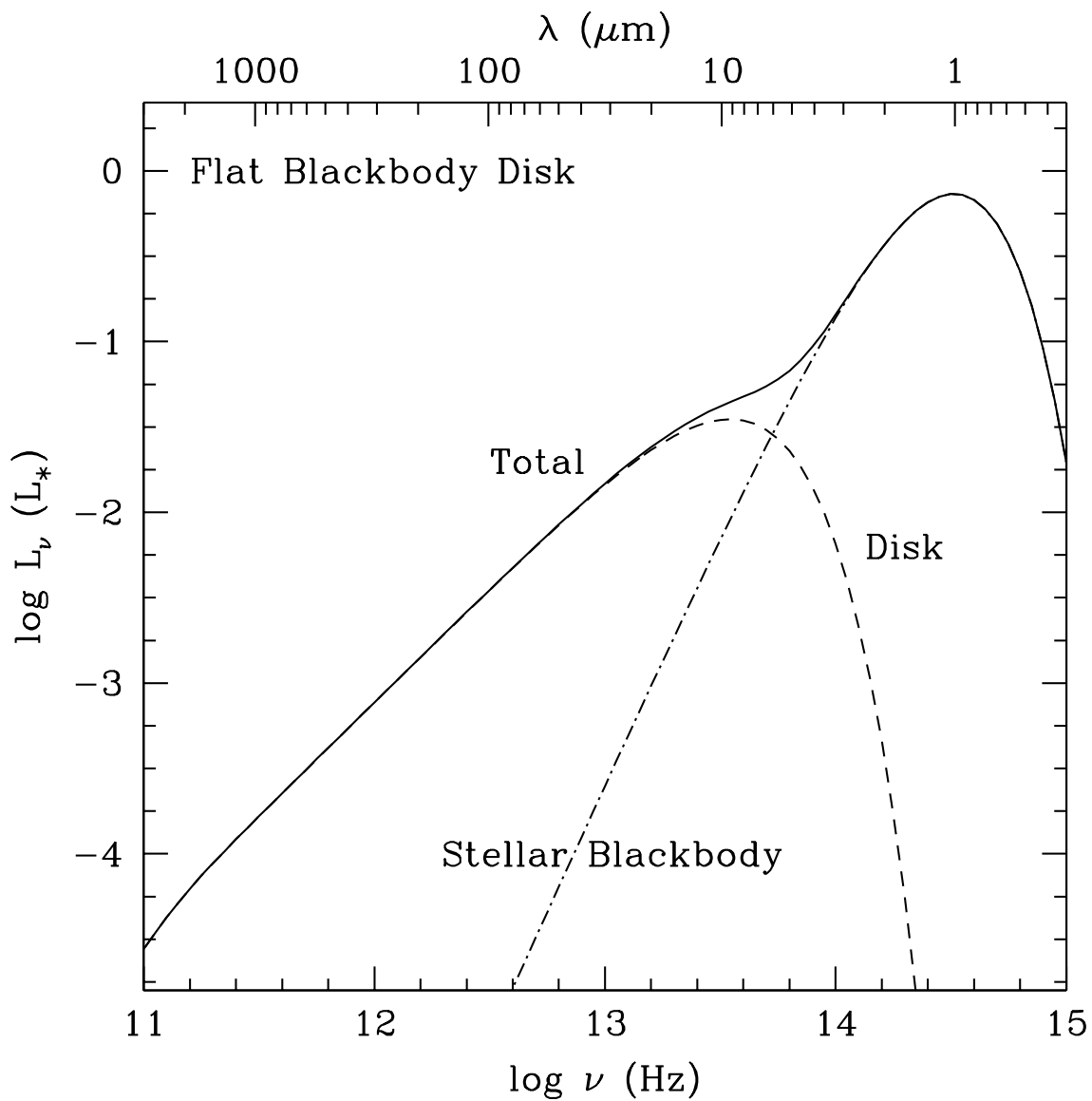}
\caption{SED for the flat blackbody disk, with contributions from star and disk
identified. The $n = 4/3$ law is evident between $30\mum$ and $1\mm$. The
turnover near $1\mm$ is due to our truncation of the disk at $a_o \approx
270\AU.$ \label{bbflatfig}}
\end{figure}

\noindent Application of the scaling relation given by equation
(\ref{sedapprox}) to the flat disk gives $L_{\nu} \sim 0.01 (\nu / 10^{13}
\Hz)^{4/3} L_*$. Figure \ref{bbflatfig} confirms that $L_{\nu}$ obeys this
relation
for $30\lesssim\lambda\lesssim 1000\mum$.

\placefigure{bbflatfig}

\subsubsection{Hydrostatic Equilibrium}
\label{bbhydro}

We retain the assumption that the disk radiates as a blackbody
and consider the consequences of vertical hydrostatic equilibrium in
a gravitational field $g=\Omega^2 z$. This is the case investigated
by \markcite{kh87}Kenyon \& Hartmann (1987).

The disk temperature $T_e$ is still given by equation (\ref{Te}),
but now the grazing angle $\alpha$ takes the more general form

\begin{equation}
\alpha\approx \frac{0.4R_*}{a}+a\frac{d}{da}\left(\frac{H}{a}\right)\, ,
\label{alpha}
\end{equation}

\noindent where $H$ is the height of the visible photosphere above
the disk midplane.

Taking the gas to be isothermal at temperature $T_e$ yields a Gaussian
vertical density profile

\begin{equation}
\frac{n}{n_0}=\exp{\left(-\frac{z^2}{2h^2}\right)}\, ,\label{denpro}
\end{equation}

\noindent where

\begin{equation}
\frac{h}{a}=\left(\frac{T_e}{T_c}\right)^{1/2}\left(\frac{a}{R_*}\right)^{1/2}\, .
\label{h}
\end{equation}

\noindent The temperature $T_c$ is a measure of the gravitational potential
at the surface of the central star;

\begin{equation}
T_c\equiv \frac{GM_*\mu_g}{kR_*}\approx 8 \times 10^6\K\,  , \label{Tc}
\end{equation}

\noindent with $\mu_g$ the mean molecular weight of the gas. Under the
assumption that the dust to gas ratio is uniform throughout the disk, we find

\begin{equation}
\frac{H}{h}=\left[2\ln\left(\frac{n_0}{n_{ph}}\right)\right]^{1/2}\, ,
\label{H/h}
\end{equation}

\noindent where $n_{ph}$ is the number density in the photosphere.

Armed with equations (\ref{alpha})--(\ref{H/h}), we are ready to evaluate
$H/a$.
In the limit of large radius where $\alpha$ is dominated by the flaring term,

\begin{equation}
\frac{H}{a}\approx 4\left(\frac{T_*}{T_c}\right)^{4/7}
\left(\frac{a}{R_*}\right)^{2/7}\approx 0.18a_\au^{2/7}\, .\label{H/a}
\end{equation}

\noindent In writing equation (\ref{H/a}) we have set $H/h = 4$; in reality,
this factor declines from about 5 at $a_\au=3$ to 4 at $a_\au=10^2$.

It follows from equations (\ref{alpha}) and (\ref{H/a}) that $\alpha\approx
0.005a_\au^{-1} +0.05a_\au^{2/7}$. Thus $\alpha$ is minimal at the transition
radius $a_{tr}\approx 0.4\au$ with $\alpha_{min}\approx 0.05$.  Beyond the
transition radius the disk flares until $H\approx a$ at $a_o \approx 270 \AU$.

\begin{figure}
\plotone{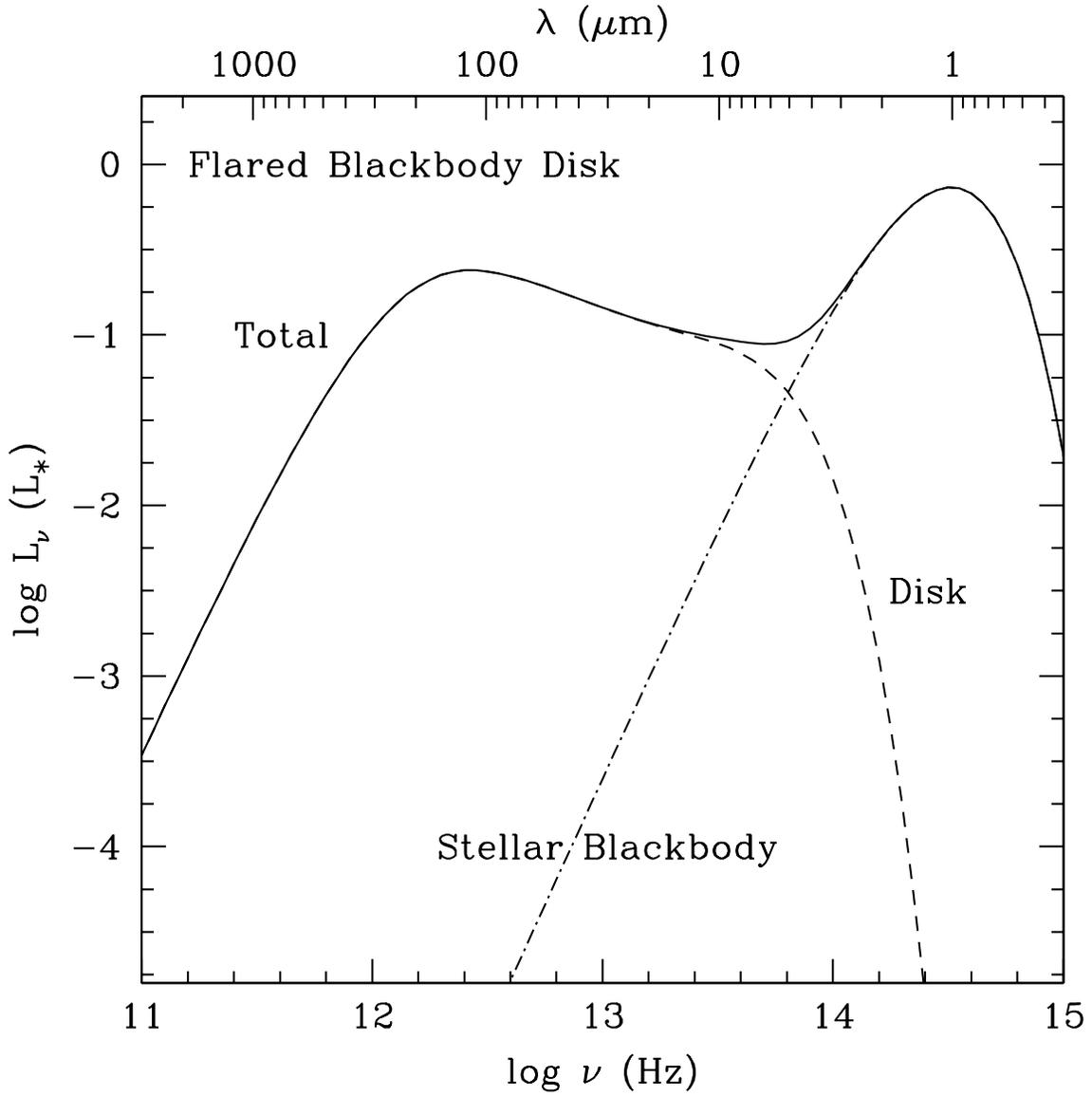}
\caption{SED for the flared blackbody disk. At mid-IR wavelengths, $L_{\nu}
\propto \nu^{-2/3}$. At longer wavelengths, $L_{\nu} \propto \nu^3$.
\label{bbflare}}
\end{figure}

The SED for the flared blackbody disk truncated at $a_o$, as computed
from equation (\ref{bbsed}), is shown in Figure \ref{bbflare}.
At wavelengths between 10 and $100 \mum$, the disk emission follows
the scaling relation [cf. eq. (\ref{sedapprox})]
$L_{\nu} \sim \alpha L_* \sim 0.1(\nu / 10^{13} \Hz )^{-2/3}L_*$.
Longward of $300 \mum$ in the Rayleigh-Jeans regime, the SED varies as $\nu^3$.

\placefigure{bbflare}

\subsection{The Radiative Equilibrium Disk}

Here we drop the blackbody assumption and determine the SED by application
of the techniques of radiative transfer. The definition of the effective
temperature, $T_e$, remains as given by equation (\ref{Te}). A simplified
description of the disk distinguishes two regions. The surface layer contains
grains which are directly exposed to light from the central star. Variables
evaluated there are denoted by a subscript $s$. We allow for the possibility
that the gas temperature in this layer, $T_{gs}$, may be smaller than the
dust temperature, $T_{ds}$. The rest of the disk
is lumped together as the interior. Variables evaluated there
carry a subscript $i$. We assume that in steady-state the interior gas and dust
temperatures are equal; $T_{gi}=T_{di}=T_i$.

\begin{figure}
\plotone{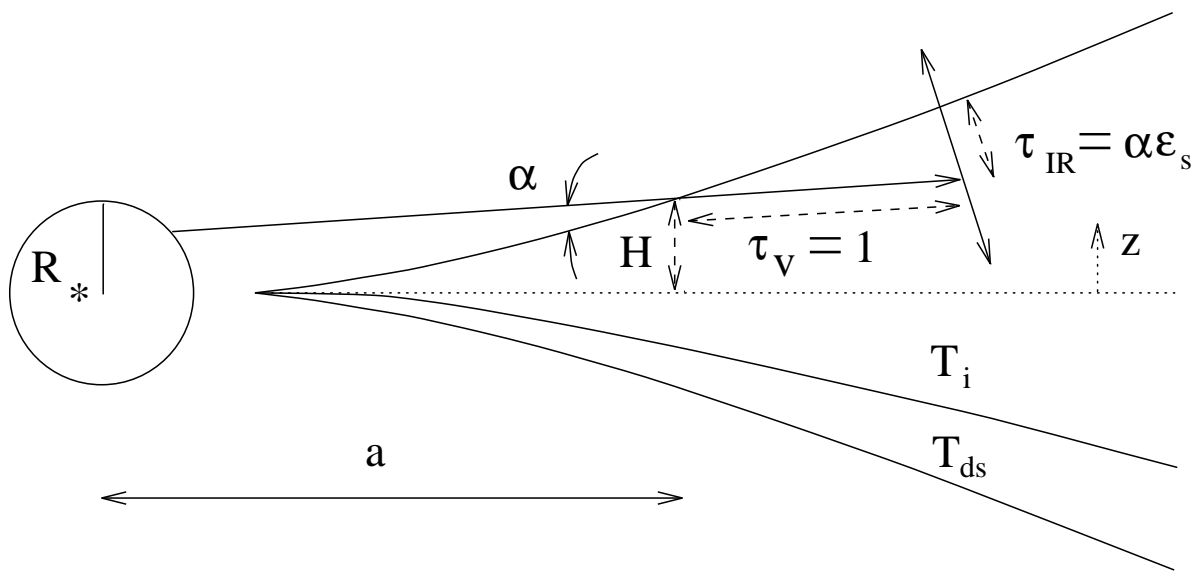}
\caption{Radiative transfer in the passive disk. Stellar radiation strikes the
surface at an angle $\alpha$ and is absorbed within visible optical depth
unity.
Dust particles in this first absorption layer are superheated to a temperature
$T_{ds}$.
About half of the emission from the superheated layer emerges as dilute
blackbody radiation.
The remaining half heats the interior to a temperature $T_i$. \label{schem}}
\end{figure}

\placefigure{schem}

A schematic of how the stellar radiation is
reprocessed is illustrated in Figure \ref{schem}. The radiation penetrates
the disk to an optical depth of order unity along
a slant path inclined by an angle $\alpha$ to the surface.
Dust particles along the slant path are ``superheated'' to a temperature

\begin{equation}
T_{ds} \approx \frac{1}{\varepsilon_s^{1/4}}\left( \frac{R_{\ast}}{2a}
\right)^{1/2} T_{\ast} \approx \frac{550}{a_\au^{2/5}} \K\,,
\label{Td}
\end{equation}

\noindent which is greater than the blackbody temperature $T_{BB}\approx
(R_*/a)^{1/2} T_*$ because their
absorptivity in the visible exceeds their emissivity in the infrared.
The normal optical depth of the superheated dust layer at visual wavelengths is
$\tau_V\approx \alpha$. The infrared optical depth is smaller still;
$\tau_{IR}\approx \alpha\varepsilon_s$.

The superheated dust radiates equal amounts of IR radiation into the inward and
outward hemispheres. The outward directed radiation is similar to that
of a dilute blackbody. Where the disk is opaque to blackbody
radiation at

\begin{mathletters}
\begin{equation}
T_{i}\approx \frac{T_e}{2^{1/4}}\approx \left(\frac{\alpha}{4}\right)^{1/4}
\left(\frac{R_*}{a}\right)^{1/2}T_* \,, \label{Ti1}
\end{equation}

\noindent the inward directed radiation is thermalized at that temperature.
The outer boundary of this region is denoted by $a_{th}$.
Just outside $a_{th}$,
the interior is optically thin to its own radiation but still opaque
to radiation from the superheated surface; in these regions the interior
temperature is determined by thermal balance to be

\begin{equation}
T_{i} \approx \left(\frac{\alpha}{4\varepsilon_i\kappa_V\Sigma}\right)^{1/4} \;
\left(\frac{R_*}{a}\right)^{1/2}\; T_* \; . \label{Ti2}
\end{equation}

\noindent At still greater radii, the encased material is transparent
to both its own radiation and to radiation from the surface; the internal
temperature here is given by

\begin{equation}
T_{i} \approx \left(\frac{\alpha \varepsilon_s^2}{\varepsilon_i}\right)^{1/4}\;
T_{ds}\approx \left(\frac{\alpha \varepsilon_s}{4\varepsilon_i}\right)^{1/4}\;
\left(\frac{R_*}{a}\right)^{1/2}\; T_*. \label{Ti3}
\end{equation}
\end{mathletters}

The SED for the radiative equilibrium disk is computed from

\begin{equation}
L_{\nu} = 8\pi^2\nu \int_{a_i}^{a_o}\,da\, a\, \int_{-\infty}^{\infty}\, dz\,
\frac{d\tau_\nu}{dz}\,
e^{-\tau_\nu}B_{\nu}(T)\; , \label{sed}
\end{equation}

\noindent where $\tau_\nu$ measures optical depth from $z$ to $\infty$
along the axis perpendicular to the disk midplane.

\subsubsection{Flat Geometry}
\label{redflat}

\begin{figure}
\plotone{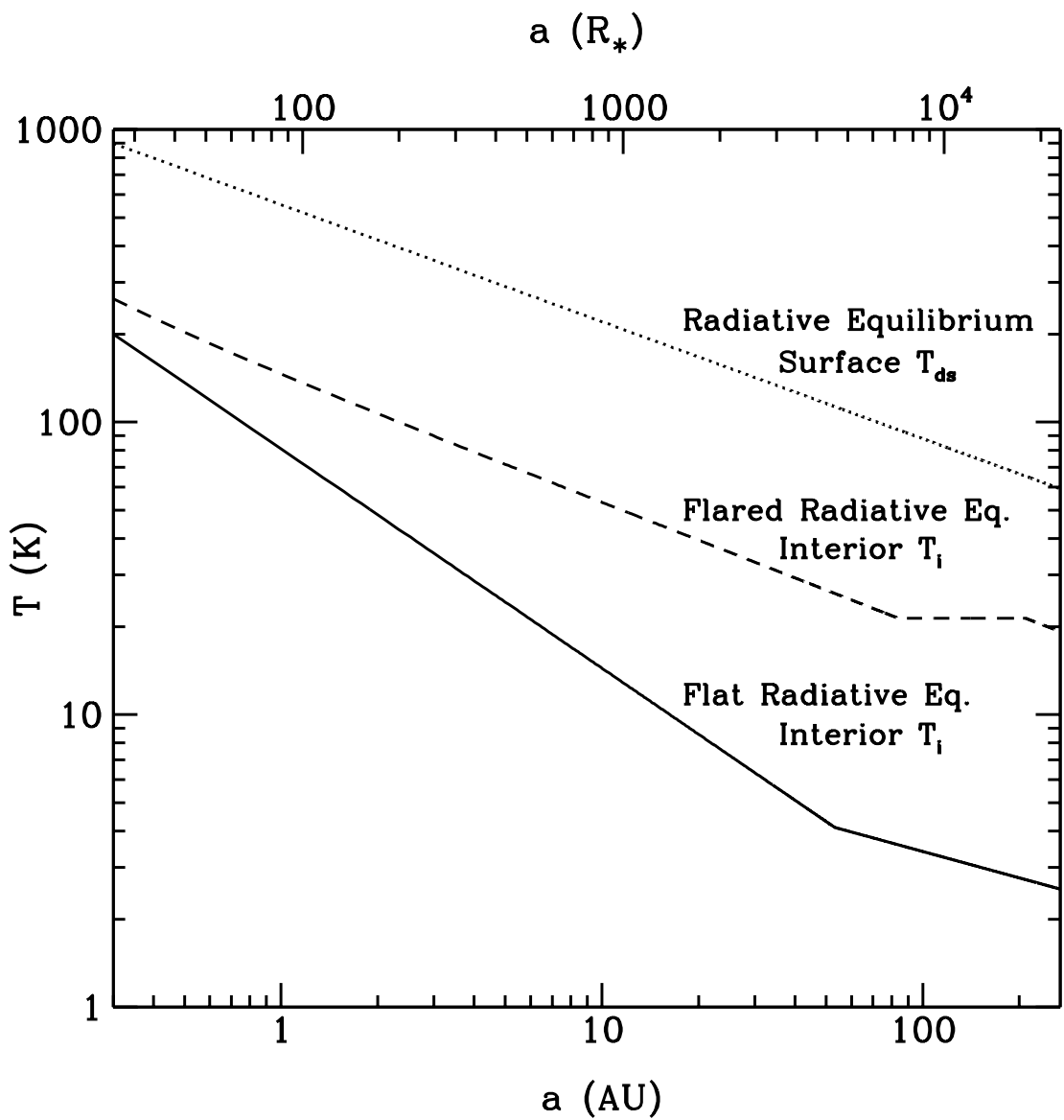}
\caption{Temperature profiles of the flat and flared radiative equilibrium
models. The dust temperature $T_{ds}$ of the superheated layer is independent
of disk geometry.
Expressions for $T_{ds}$ and $T_i$ are provided in the text.
The flat disk is truncated at $a_o = 270\AU$ (to facilitate comparison
with the flared models), before the third temperature regime is reached.
\label{tprof}}
\end{figure}

Once again we consider the flat disk, but now under conditions of radiative
equilibrium. The appropriate expression for the effective temperature is given
by equation (\ref{Tef}). For our fiducial flat disk $a_{th} \approx 50 \AU$.
Runs of $T_{ds}$ and $T_i$ as functions of $a$ are displayed in Figure
\ref{tprof}.
\placefigure{tprof}

\begin{figure}
\plotone{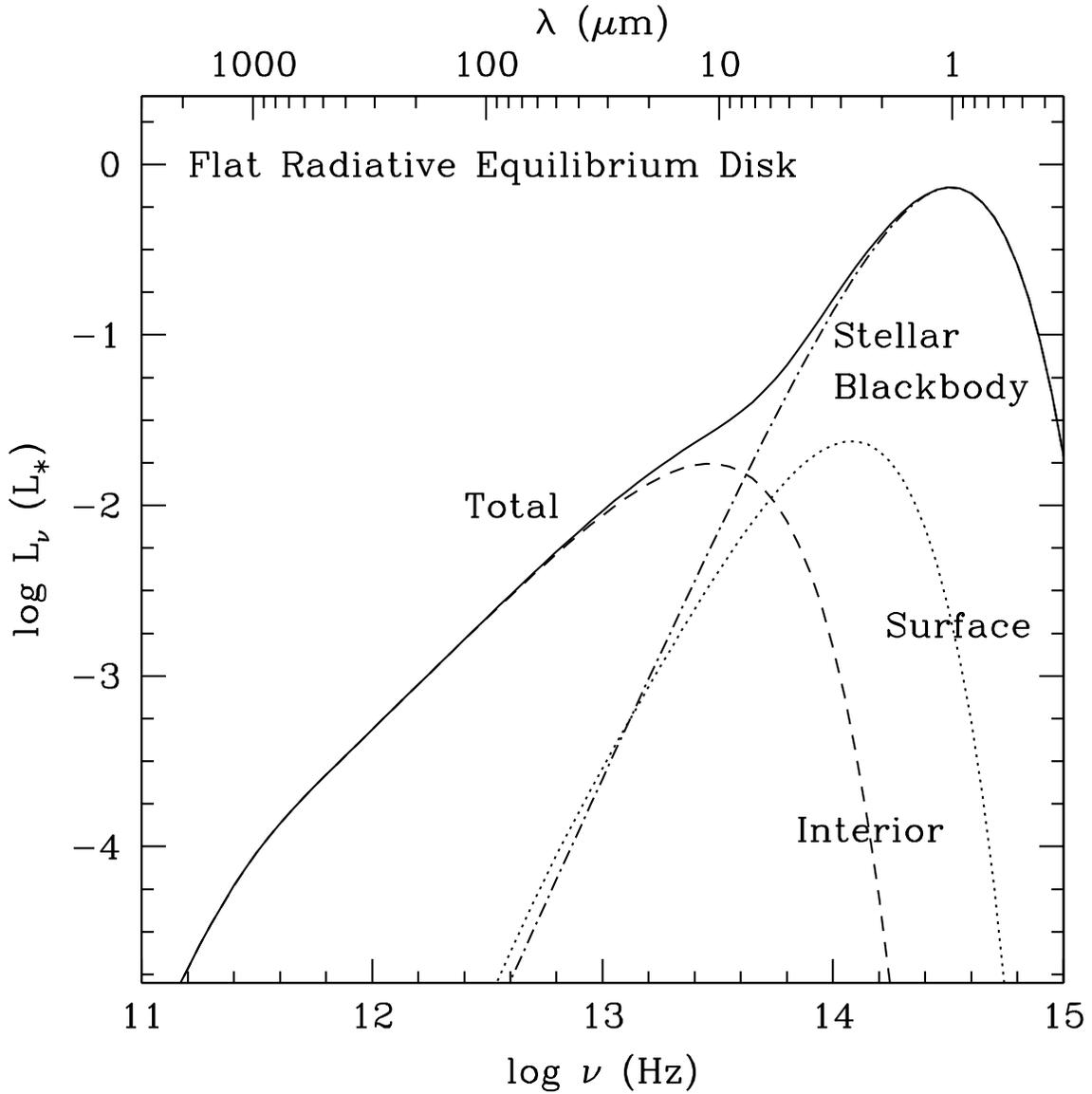}
\caption{SED for the flat radiative equilibrium disk. Emission from the
superheated surface is hidden by central starlight. For $30 \lesssim \lambda
(\mum) \lesssim 1000$,
optically thick emission from the disk interior resembles that from the flat
blackbody disk
(see Figure \protect{\ref{bbflatfig}}), but is reduced by a factor of 2.
Emission longward of $1 \mm$ is optically thin. \label{reflat}}
\end{figure}

The SED for the flat, radiative equilibrium disk as calculated from equation
(\ref{sed}) is displayed in Figure \ref{reflat}. Its appearance is similar to
that of the SED for the flat
blackbody disk. Over most of the IR, it is dominated by radiation
from the optically thick interior. The surface layer
radiates more than the interior shortward of $6\mum$; however,
there its contribution is hidden by that from the
central star. Most of the radiation longward of a millimeter comes from
the outer, optically thin part of the disk. This accounts for the drop of
the SED below the extrapolation of the $n = 4/3$ power law.

\placefigure{reflat}

\subsubsection{Hydrostatic Equilibrium}
\label{redhydro}

Now we investigate the disk model in which both vertical
hydrostatic equilibrium and radiative transfer are treated in a
self-consistent fashion. The flaring geometry is governed by equations
(\ref{alpha})
through (\ref{H/h}). In the limit where the grazing angle is dominated
by the flaring term and the disk is opaque to its thermal radiation,
the expression for $H/a$ is nearly identical to that given by
equation (\ref{H/a}). In other, more optically thin regimes, the
flaring geometry changes slightly as $T_i$ takes successively different
forms.

A plot of $T_i$ for the flared, radiative
equilibrium disk is included in Figure \ref{tprof}.
Approximate fitting formulae for $T_i$ and $H/a$ in the three regions
of the flared, radiative equilibrium disk are as follows.
For $0.4 < a_\au < 84$,

\begin{mathletters}
\begin{equation}
T_{i} \approx  \frac{150}{a_\au^{3/7}} \,\K
\end{equation}
\begin{equation}
H/a \approx  0.17 \,a_\au^{2/7} \,;
\end{equation}

\noindent for $84 < a_\au < 209$,

\begin{equation}
T_{i} \approx  21\,\K
\end{equation}
\begin{equation}
H/a \approx  0.59 \,\left(\frac{a_\au}{84}\right)^{1/2} \,;
\end{equation}

\noindent and for $209 < a_\au < 270$,

\begin{equation}
T_{i} \approx  21 \,\left(\frac{209}{a_\au}\right)^{19/45} \,\K \\
\end{equation}
\begin{equation}
H/a \approx  0.92 \,\left(\frac{a_\au}{209}\right)^{13/45} \,.
\end{equation}
\end{mathletters}

\begin{figure}
\plotone{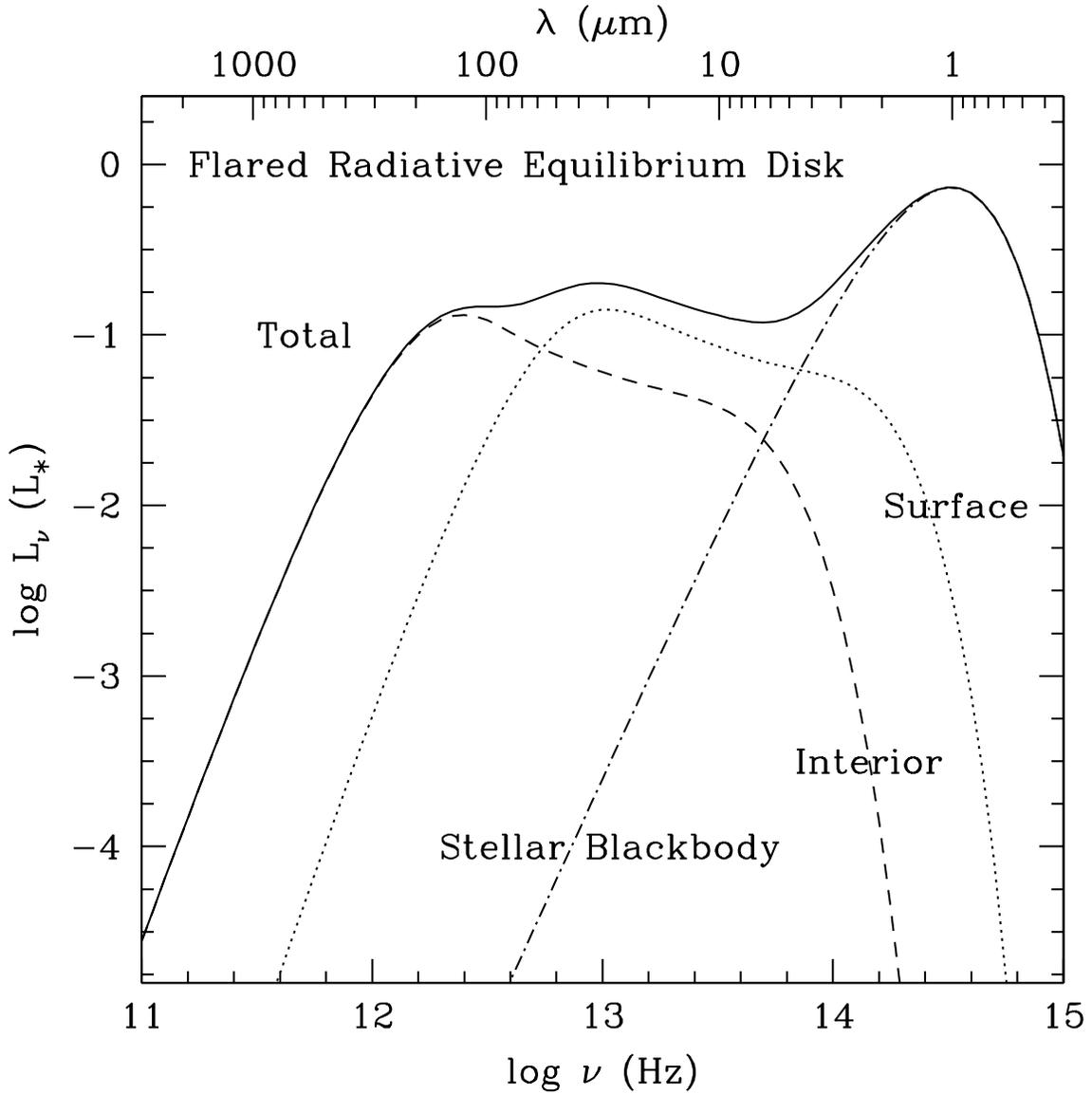}
\caption{SED for the hydrostatic, radiative equilibrium disk. At mid-IR
wavelengths, the superheated surface radiates approximately 2--3 times more
power than the interior.
Longward of $300 \mum$, $n$ gradually steepens from about 3 to
$3 +\beta$ as the disk becomes increasingly optically thin.\label{reflare}}
\end{figure}

The SED for the flared disk truncated at $a_o \approx 270 \AU$, as
computed from equation (\ref{sed}), is shown in Figure
\ref{reflare}.\footnote{The
disk may terminate in a slow thermal wind since $H/a\approx 1$ at
$a_\au=270$.} Radiation emitted by the surface mirrors
that of the interior, but is shifted towards and dominates at shorter
wavelengths. The near power law rises of both disk components of the
SED in the mid-infrared are readily obtained following the procedure used in
the derivation of equation (\ref{sedapprox}). The superheated layer
obeys the scaling relation $L_{\nu, s} \sim 8\pi a^2
\tau_{IR}\;\sigma T_{ds}^4 \sim \alpha L_*/2 \sim 0.06 (\nu /10^{13}
\Hz)^{-5/7} L_*$.  Similarly, the opaque interior obeys $L_{\nu, i} \sim
8\pi  a^2 \sigma T_i^4\sim \alpha L_*/2 \sim 0.02 (\nu / 10^{13} \Hz)^{-2/3}
L_*$.  That the former exceeds the
latter at fixed wavelength reflects the fact
that compared to $L_{\nu, i}$, most of the contribution to $L_{\nu, s}$ arises
from larger radii and hence larger $\alpha$.

\placefigure{reflare}

\section{EXTENSIONS AND REFINEMENTS}

\subsection{The Planetary Atmosphere Analogy}
\label{opac}

Planetary atmospheres are generally transparent to visible radiation
above their uppermost clouds decks.\footnote{Column densities above the cloud
tops are similar to that of
the minimum mass solar nebula at $a_\au\approx 1$, $\Sigma\sim
10^3\gm\cm^{-2}$.} Visible radiation is absorbed by particles.
Collisions maintain thermal equilibrium between gas molecules and the
particles. Unit optical depth in the thermal infrared is reached
above the cloud tops. Pressure broadened molecular lines are
the main source of infrared opacity.\footnote{Collision induced dipole
transitions of molecular hydrogen dominate in the atmospheres of the
outer planets.} The total infrared luminosity, the sum of absorbed
sunlight and internal heat loss, is emitted as approximate black body
radiation. Thus the energy balance of a planetary atmosphere resembles
that of the blackbody disk.\footnote{The analogy even extends to the
association of the planet's net luminosity with the accretion
luminosity of the disk.} This analogy probably accounts for
the casual acceptance of the blackbody disk model.

However compelling, this analogy is flawed. T Tauri disks differ from
planetary atmospheres in a crucial parameter, the vertical
gravitational acceleration.  For disks, $g_{disk}\approx \Omega^2\,z$,
with $\Omega$ the orbital angular velocity, and $z$ the distance
from the midplane. For our standard parameters, $g_{disk}\approx
0.4\, z_\au\, a_\au^{-3}\cm\s^{-2}$, much smaller than $g_{planet}\sim
10^3\cm\s^{-2}$ for $a \gtrsim a_i$.
Three consequences of the low vertical gravitational acceleration in disks
which invalidate the analogy with
planetary atmospheres are described below.

Dust grains settle slowly in disks. The gravitational settling time to the
midplane,

\begin{equation}
t_{settle}\approx \frac{\Sigma}{r\rho_d\Omega}\approx
10^7\left(\frac{0.1\mum}{r}\right)\left(\frac{\Sigma_0}{10^3\gm\cm^{-2}}\right)
\yr\, ,\label{tsettle}
\end{equation}

\noindent is independent of $a$. It is longer than the disk lifetime for
particles which most efficiently absorb stellar radiation. Moreover, the
settling may be slowed by turbulent mixing or vertical circulation of the gas.

The gas pressure in disks is so low that dust grains which are
directly exposed to stellar radiation reradiate almost all of
the stellar energy they absorb. Thus the temperature of the
superheated grains is determined by radiative balance and is independent of the
ambient gas temperature. The regulation of the gas temperature is a more subtle
issue (cf. \S \ref{ebal}).

The low pressure in disks also means that molecular lines suffer negligible
pressure broadening. Thus the gas opacity
is concentrated in narrow, Doppler broadened lines.
Important molecular sources of opacity must have permitted dipole transitions
near the peak of the blackbody spectrum corresponding to the ambient
temperature $T$. Furthermore, these transitions must connect energy levels
which do not lie too far above $kT$. Water is a prime candidate
for a coolant at temperatures greater than $100\K$. It is abundant, and as a
consequence of being both a hydride and an asymmetric rotor, it
possesses a rich rotation spectrum shortward of $300\mum$. However, our
crude estimates suggest that even if much of the cosmic abundance of oxygen is
tied up in water, water lines would not cover more than a small
fraction of the infrared spectrum at any temperature.

\subsection{Energy Balance In The Superthermal Dust Layer}
\label{ebal}

Stellar energy absorbed by dust grains is lost through
emission of infrared radiation and collisional transfer to gas molecules. The
gas molecules, mostly H$_2$ and He, gain energy in collisions with dust grains
and lose it in collisions with molecular coolants such as H$_2$O. The energy
added to the internal degrees of freedom of the
coolants in collisions with gas molecules and
by absorption of dust radiation is lost in radiation to
space. Thus $T_{ds}\gtrsim T_{gs}\gtrsim T_x$, where the subscripts
$j = \{ds,gs,x\} $ denote quantities pertaining to the dust, the inert gas
molecules, and the molecular coolants, respectively. \footnote{The symbol $T_x$
refers to the excitation temperature of the coolant transition.}

\begin{figure}
\plotone{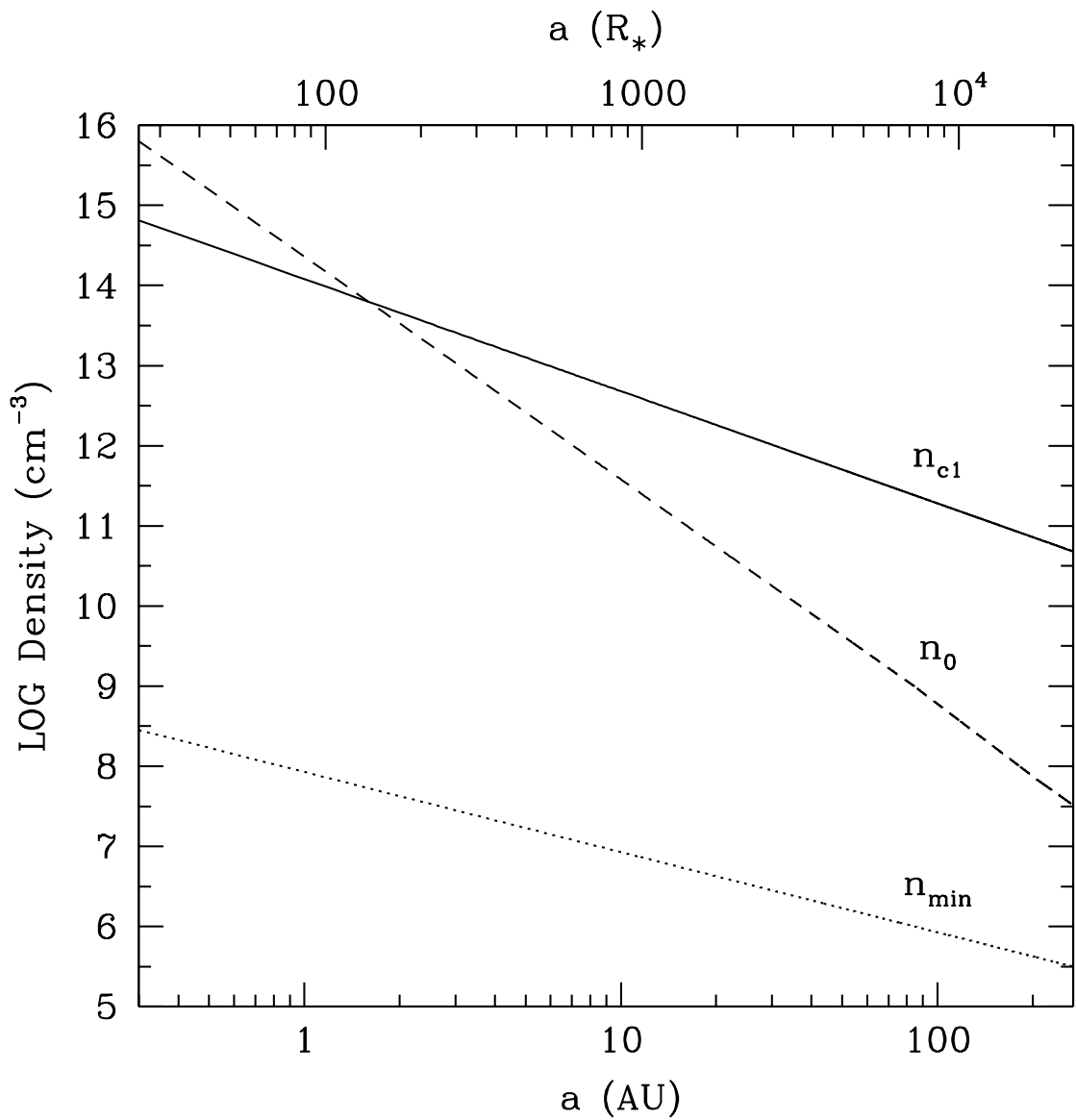}
\caption{Comparison of possible gas densities in the superheated
layer to the critical density $n_{c1}$ above which gas-grain
collisions might lower the dust temperature. The actual gas density could be as
high as $n_0$, the value at the disk midplane, or as low as $n_{min}$,
the value for a disk in which gas and dust are uniformly mixed everywhere.
Except in
the innermost $2\au$, it is evident that the dust likely remains superheated at
the
temperature given equation (\protect{\ref{Td}}). \label{nprof}}
\end{figure}

Balancing the rates of energy gain and loss per unit surface area of a dust
grain yields

\begin{equation}
\sigma T_*^4 \left(\frac{R_*}{a}\right)^2 \approx  \varepsilon_s\sigma T_{ds}^4
 + n_g v_g k(T_{ds} - T_{gs}) \label{dbal}\, ,
\end{equation}

\noindent with $v_g$ the sound speed.  For $n_g\ll n_{c1}$, where the critical
number density

\begin{equation}
n_{c1}\approx \frac{\varepsilon_s\sigma T_{ds}^3}{v_g k}\approx
\frac{\sigma\mu_g^{1/2}T_*^{5/2}}{k^{3/2}}
 \left(\frac{R_*}{a}\right)^{7/5}\approx \frac{10^{14}}{a_\au^{7/5}}\cm^{-3}\,
, \label{nc1}
\end{equation}

\noindent gas-grain collisions have a negligible effect on $T_{ds}$ and
equation (\ref{dbal}) reduces to equation (\ref{Td}).\footnote{We evaluate
$v_g$ on the right hand side of equation (\ref{nc1}) at temperature $T_{ds}$.}
On the other hand, for $n_g\gg n_{c1}$, $T_{ds}\approx T_{gs}$. The
gas density in the superheated dust layer, $n_{gs}$, is unknown. It
could be as large as $n_0$, the value at the disk midplane, or as low
as $n_{min}$, the value obtained by assuming uniform mixing of dust and
gas up to the tenuous outer layers of the disk. Runs of $n_{c1}$, $n_0$,
and $n_{min}$ as functions of $a$ are displayed in Figure \ref{nprof}. It is
seen that even $n_0$ falls below $n_{c1}$ beyond $a_\au\approx 2$.

\placefigure{nprof}

Balancing the rates at which inert gas molecules gain and lose energy in
collisions with dust grains and molecular coolants yields

\begin{equation}
n_{ds} \sigma _d (T_{ds} - T_{gs}) \approx n_x \sigma _x (T_{gs} - T_x)
\label{gbal}\, ,
\end{equation}

\noindent where $n$ is a number density, and $\sigma$ denotes a
collisional cross-section. The compositions and mixing ratios of
molecular coolants are uncertain.  Photodissociation and incorporation
into dust grains represent loss mechanisms that can reduce abundance
ratios far below values based on cosmic abundances.  Nevertheless, suppose
that the mass in coolant molecules is comparable to that in dust
grains. Then $n_{ds}\sigma_d/n_x\sigma_x\sim 10^{-3}$, the ratio of the
linear size of a gas molecule to a dust grain. Thus we would not be
surprised to discover that $T_{gs}-T_x\ll T_{ds}-T_{gs}$.

Estimating excitation temperatures for the transitions in molecular coolants is
a daunting task. We proceed by making a number of drastic approximations.
Each transition is assumed to give rise to an optically thick, Doppler
broadened line having excitation temperature $T_x$. For clarity, we
begin by considering the pure radiative excitation of these levels. It is not
difficult to show that the excitation comes mainly from radiation emitted by
the superheated dust grains as opposed to that from the
inner portion of the disk. We specialize further to transitions
whose frequencies lie on the Rayleigh-Jeans part of the blackbody spectrum at
temperature $T_{ds}$. Applying the basic concepts of the escape probability
formalism for radiative transfer to a two level molecule, we obtain

\begin{mathletters}
\begin{eqnarray}
T_x\approx \alpha\varepsilon_s T_{ds}\quad &{\rm for} \quad h\nu\lesssim
kT_x\,, \\
T_x\approx \frac{h\nu}{k}\ln\left(\frac{1}{\alpha\varepsilon_s}
\frac{h\nu}{kT_{ds}}\right)\quad & {\rm for} \quad h\nu\gtrsim kT_x\, .
\end{eqnarray}
\end{mathletters}

\noindent Pure radiative excitation sets a floor on the value of
$T_x$.

Next we consider the effects of collisions while neglecting
radiative excitation.  Here we suppose that lines of brightness
temperature $T_x$ cover a fraction $f$ of the blackbody
spectrum at temperature $T_x$. We equate the emission line flux
to the rate (per unit disk surface area) at which the
gas gains energy in collisions with the superheated dust grains:

\begin{equation}
f\sigma T_x^4 \approx n_{gs}v_gk\alpha (T_{ds}-T_{gs})\, .
\label{cbal}
\end{equation}

\noindent For $n_{gs} \ll n_{c2}$, where the second critical gas
density

\begin{equation}
n_{c2} \approx \frac{f}{\alpha \varepsilon_s} n_{c1}\,,
\label{nc2}
\end{equation}

\noindent we find

\begin{mathletters}
\begin{equation}
T_x\approx \left(\frac{n_{gs}}{n_{c2}}\right)^{1/4}T_{ds}\, .
\label{txlow}
\end{equation}

\noindent In the high density limit $n_g \gg n_{c1}$,
$T_x \approx T_{gs} \approx T_{ds} \approx T_s$ with

\begin{equation}
T_s\approx \left(\frac{\alpha}{f+\alpha
\varepsilon_s}\right)^{1/4}\left(\frac{R_*}{a}\right)^{1/2}T_* \, .
\label{onet}
\end{equation}
\end{mathletters}

\noindent If $f\ll \alpha \varepsilon_s$, equation (\ref{onet}) reduces to
(\ref{Td}), the
appropriate expression for the dust temperature in the absence of cooling by
molecular line emission.
On the other hand, if $f\gg \alpha \varepsilon_s$, (\ref{onet}) collapses
to the expression appropriate for a blackbody disk as given by equation
(\ref{Te}),
magnified by a factor of about $1/f^{1/4}$.

\subsection{Location Of The Superthermal Layer}
\label{settlesh}

We have been assuming that the dust and gas are uniformly mixed
throughout the disk. Consequently, for our model the gas density is much
smaller
in the superthermal dust layer than in
the disk midplane (cf. Fig. \ref{nprof}). Although the timescale for dust
grains to settle to the disk midplane may be long compared to the disk lifetime
[cf. equation (\ref{tsettle})], the timescale for grains to fall through the
superthermal layer (in the absence of advection by the gas) is considerably
shorter. The latter is given by

\begin{equation}
t_{settle,s} \sim \frac{v_g}{a r \kappa _V\rho_d \Omega ^2} \sim a_\au^2
\left(\frac{0.1 \mum}{r}\right) \left(\frac{400 \cm^2 \gm^{-1}}{\kappa
_V}\right) \yr \, , \label{tsettlesh}
\end{equation}

\noindent where $\kappa _V$ is the dust opacity in the surface layer at visual
wavelengths. The timescale
for dust collisions/coagulation to occur in the superthermal layer is $1/\alpha
$ times larger
than $t_{settle,s}$. Thus, the superthermal dust layer is likely to
be found closer to the disk midplane and to have a lower $\kappa_V$ than our
simple assumption of uniform dust-gas mixing implies.

How might we learn about the level at which the superthermal dust layer
sits? One possibility is through the observation of emission lines
that would form in the gas associated with this layer if the gas temperature
is comparable to the dust temperature. Perhaps most revealing would be
observations
of the rotation lines of H$_2$. Because these arise from quadrupole transitions
and are therefore weak,
they would only be seen if the dust layer were located where the gas density
is much larger than the minimum value given by the assumption of uniform
mixing of dust and gas.

\subsection{Dust Albedo}
\label{albedo}

A nonzero dust grain albedo, $A_d$, at visible wavelengths would reduce the
absorbed stellar flux by a factor of order $(1-A_d)^{1/2}$. This would lower
the disk thickness, but by a less impressive factor of order $(1-A_d)^{1/16}$.

How small might $(1-A_d)^{1/2}$ be? Values of $A_d\sim 0.4-0.5$ are
typical of interstellar grains \markcite{dl84} (Draine \& Lee 1984). Visual
albedos of
pure ice particles are much closer to unity. And ices are likely to
comprise a significant fraction of the mass in grains in the outer
disk. Starting inside $a_\au=10^2$, the partial pressures of H$_2$O,
then NH$_3$, and finally CH$_4$ are expected to exceed their vapor
pressures. However, taking the outer planet atmospheres as a guide, it seems
unlikely that any ice so formed would be pure enough to yield
$(1-A_d)^{1/2}\lesssim 0.3$.

\subsection{Accretional Heating}

A star which derives its radiant energy from accretion has $T_{\ast}
\approx (G M_*\dot{M}_*/R_*^3 \sigma)^{1/4}$, where $\dot {M}_*$ is the
mass accretion rate.\footnote{Provided the accretional energy is
radiated over the entire stellar surface.} The stellar flux impinging
on the circumstellar disk, $\sigma T_{\ast}^4 (R_{\ast}/a)^2 \alpha$,
is to be compared with the local viscous dissipation rate per unit
area, $\sigma T_{\ast}^4 (R_{\ast}/a)^3$. Thus in the extreme case that
accretion accounts for the entire stellar luminosity, viscous dissipation is
only competitive with stellar heating for $a\lesssim a_{tr}$.

The bulk of the viscous dissipation probably takes place deep inside the
disk. Provided the disk is sufficiently opaque, this could result in
midplane temperatures being higher than $T_*(R_*/a)^{3/4}$ by
a factor of order $\tau ^{1/4}$, where $\tau\approx
\varepsilon_i\kappa_V\Sigma$ is the effective vertical
optical depth. A concomitant increase in the disk thickness by a factor
of $\tau ^{1/8}$ changes the surface geometry and
may cause some portions of the disk to be shadowed
from direct exposure to starlight. Such effects are likely to
be important only for the innermost regions of the disk where radiation
temperatures, surface densities, and consequently optical depths are
highest. For example, around an accretion powered star, the midplane
temperature of our standard disk is about $2000\K$ at $a_\au = 1.5$. This
yields about a factor 2 increase in disk thickness relative to that
of the passive disk. For smaller radii, the enhanced thickness
would be buffered somewhat by the reduction of opacity associated with the
vaporization of silicates.

\section{DISCUSSION}

\subsection{Comparison with Observations}

\subsubsection{Continuum SEDs}

\begin{figure}
\plotone{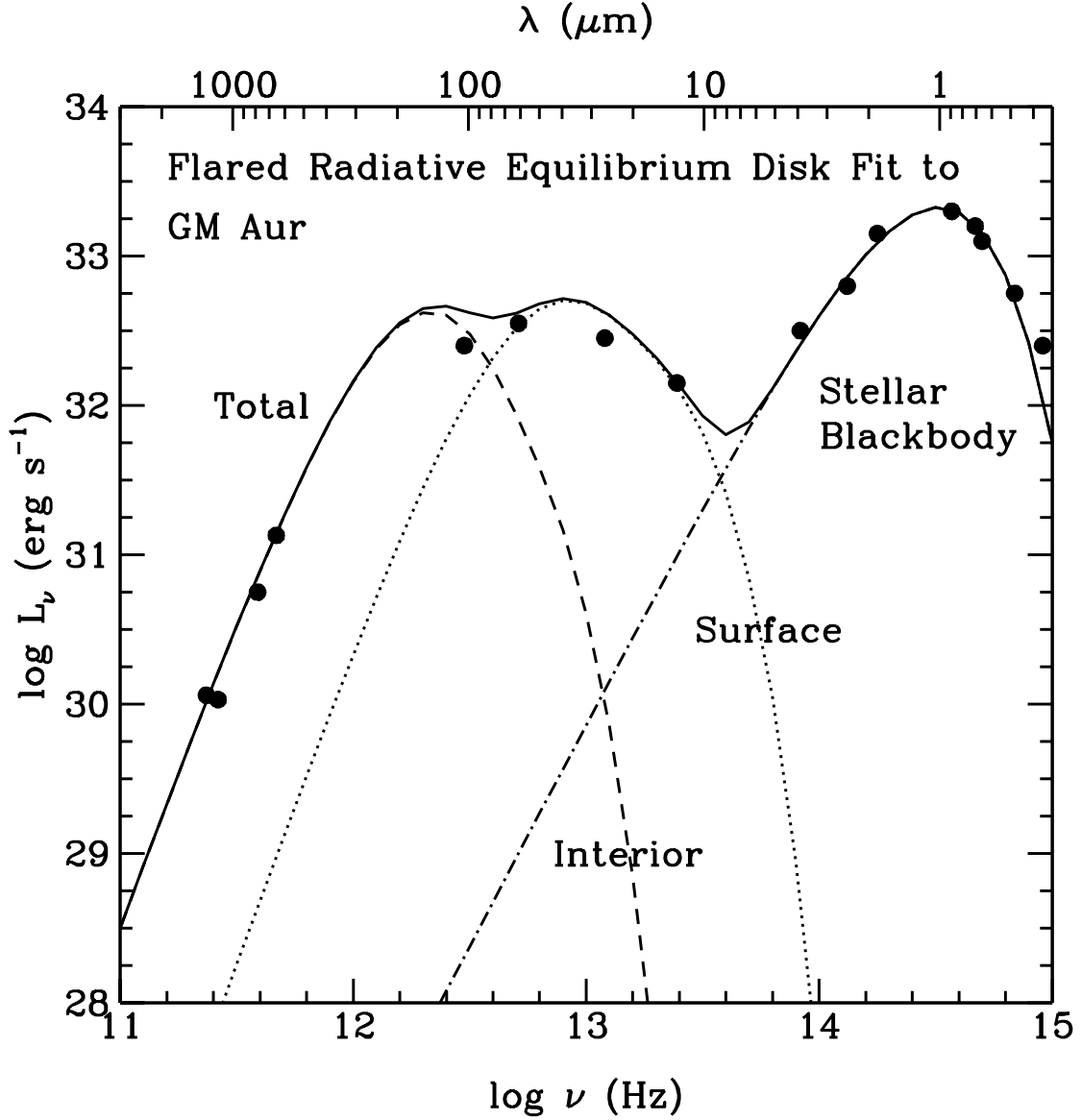}
\caption{Observed SED of GM Aur (solid points) and accompanying
hydrostatic, radiative equilibrium (face-on) disk model. Fit parameters are as
follows:
$\Sigma = 3 \times 10^3 a_\au^{-3/2} \gm \cm^{-2}$, $\beta = 1.4$, $r_p =
0.3\mum$, $a_o = 390\au$,
and $a_i = 5.7\au$. The derived inner cut-off radius is likely an artifact of
fitting a face-on model to a disk observed at substantial inclination.
See the text for a discussion. \label{gmaur}}
\end{figure}

In Figure \ref{gmaur}, we demonstrate that the nearly flat infrared
excess of the classical T Tauri star GM Aur can
be naturally interpreted as arising from a passive
reprocessing disk.
Parameters for the central star are obtained from \markcite{betal90} Beckwith
et al. (1990),
and a distance of 140 pc to the Tau-Aur cloud is assumed.
Approximate best fit parameters for the assumed face-on disk are the same as
those of
our fiducial model, with the following alterations:
$\beta = 1.4$, $r_p = 0.3\mum$, $\Sigma_0 = 3 \times 10^3\gm\cm^{-2}$, $a_o = 4
\times 10^4 R_* = 390\AU$, and $a_i = 680 R_* = 5.7\AU$.

\placefigure{gmaur}

The inner cut-off radius is chosen so that the model matches the
observationally
determined SED at $12\mum$; a smaller cut-off radius results in excessive
emission. However, it seems likely that the central hole is an artifact of
fitting a face-on model to observations of a disk viewed at a non-zero
inclination angle, $i$.  Millimeter-wave aperture synthesis images in CO
and Hubble Space Telescope (HST) images in scattered stellar light reveal
a flared circumstellar disk in Keplerian rotation \markcite{k97}
(Koerner 1997), and suggest that $i\sim 60^{\circ}$.
It is possible that the flared outer ``wall'' extincts emission from the
inner disk, diluting the $12\mum$ flux and mimicking the signature of an
inner hole.
We reserve a more detailed discussion of the effects of non-zero inclination
angle to a future study.

\subsubsection{Spatially Resolved Broadband Observations}
Pioneering observations which marginally resolve disks at
mm wavelengths are reported in \markcite{letal94} Lay et al. (1994) and
\markcite{metal96} Mundy et al. (1996). Fits of elliptical Gaussians to the
$\lambda = 0.87\mm$ brightness distributions of HL Tau and L1551 IRS 5
yield semi-major radii at half-maximum brightness of 60 and $80\AU$,
respectively \markcite{letal94} (Lay et al. 1994). Similar
results were obtained at $\lambda = 2.7\mm$ for HL Tau, with the semi-minor
radius marginally resolved to be $\sim30\AU$ \markcite{metal96} (Mundy et al.
1996).

\begin{figure}
\plotone{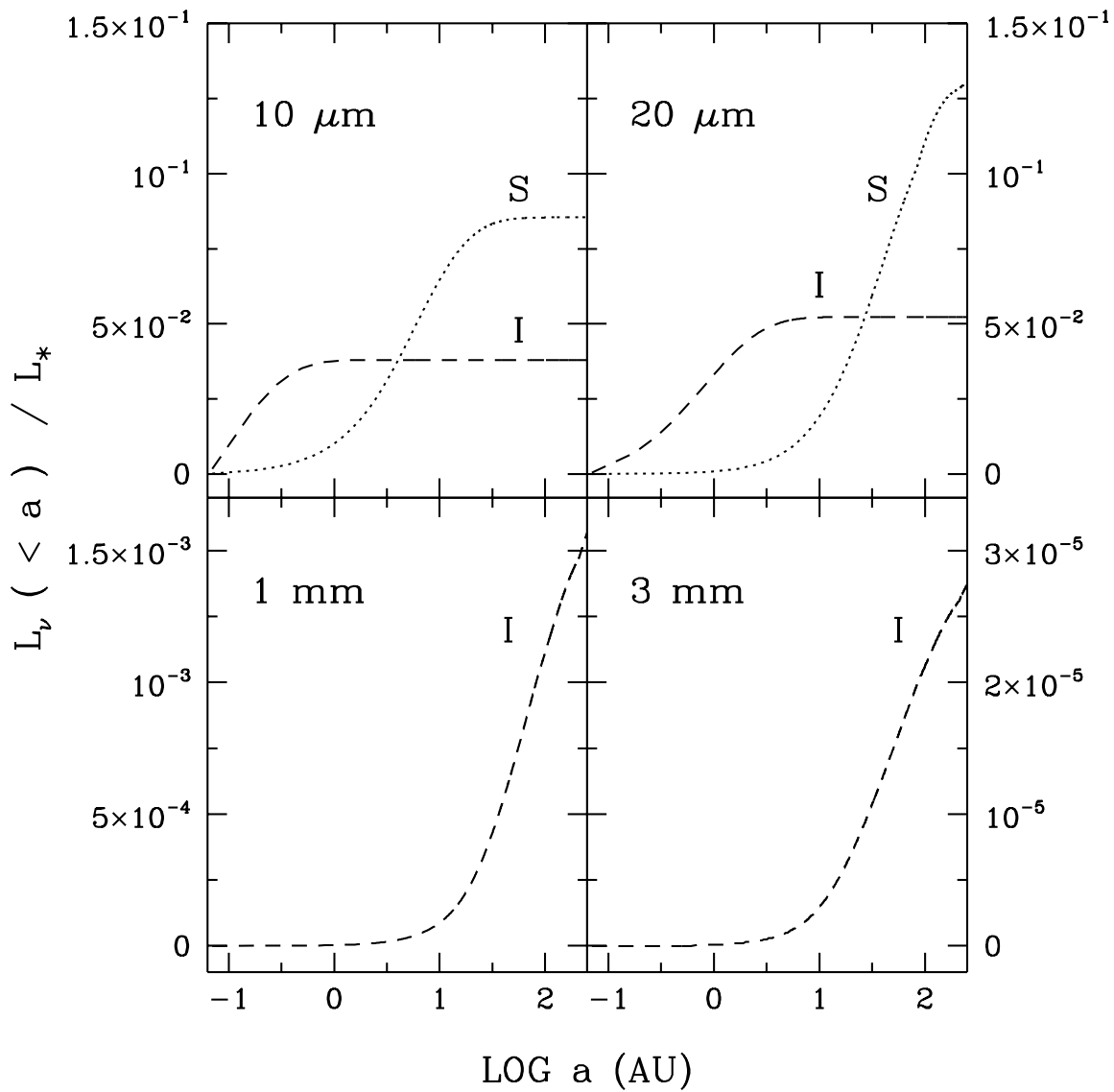}
\caption{Running contributions to the SED from the superthermal dust layer (S)
and the disk
interior (I) at wavelengths for which the disk may be increasingly resolved
by interferometers. For a given disk component, $L_{\nu} ( < a )$ is given by
equation (\protect{\ref{sed}}),
with $a_o$ replaced by $a$. At $\lambda = 1$ and $3\mm$, the contributions from
the surface layer are negligible
and not shown. \label{fluxfig}}
\end{figure}

Interferometric observations of the nearest T Tauri disks at $\lambda\approx
10\mum$ with 10 milliarcsecond resolution might separate the contributions to
the SED from the superthermal dust layer and the disk interior.
Running integrals for $L_{\nu, s}$ and $L_{\nu, i}$ against $a$ are displayed
in Figure \ref{fluxfig}.
Most of the $10\mum$ emission inwards of $1\AU$ originates from the disk
interior,
whereas the contribution from the superthermal layer is localized in an annulus
centered on $\sim 10\AU$. Similar results hold for the $20\mum$ flux.
At wavelengths longer than $\sim 100\mum$, the disk interior dominates
the emission.

\placefigure{fluxfig}

\subsubsection{Spectral Dust Features}
\label{specobs}

Dust grains possess spectral resonances. Among the best-known are those at
$\lambda=9.7\mum$ and $18\mum$ which are
thought to arise from silicates (see, e.g., \markcite{d95} Draine 1995); the
former has been observed in both
emission and absorption in the spectra of T Tauri stars. For example,
\markcite{cw85} Cohen \& Witteborn (1985)
identified $10\mum$ emission features in 24 stars
and $10\mum$ absorption features in another $7$ in a spectrophotometric survey
of 32 T Tauri stars. The emission features evince
line to continuum ratios of about 1.2--3.0 to 1.

\begin{figure}
\plotone{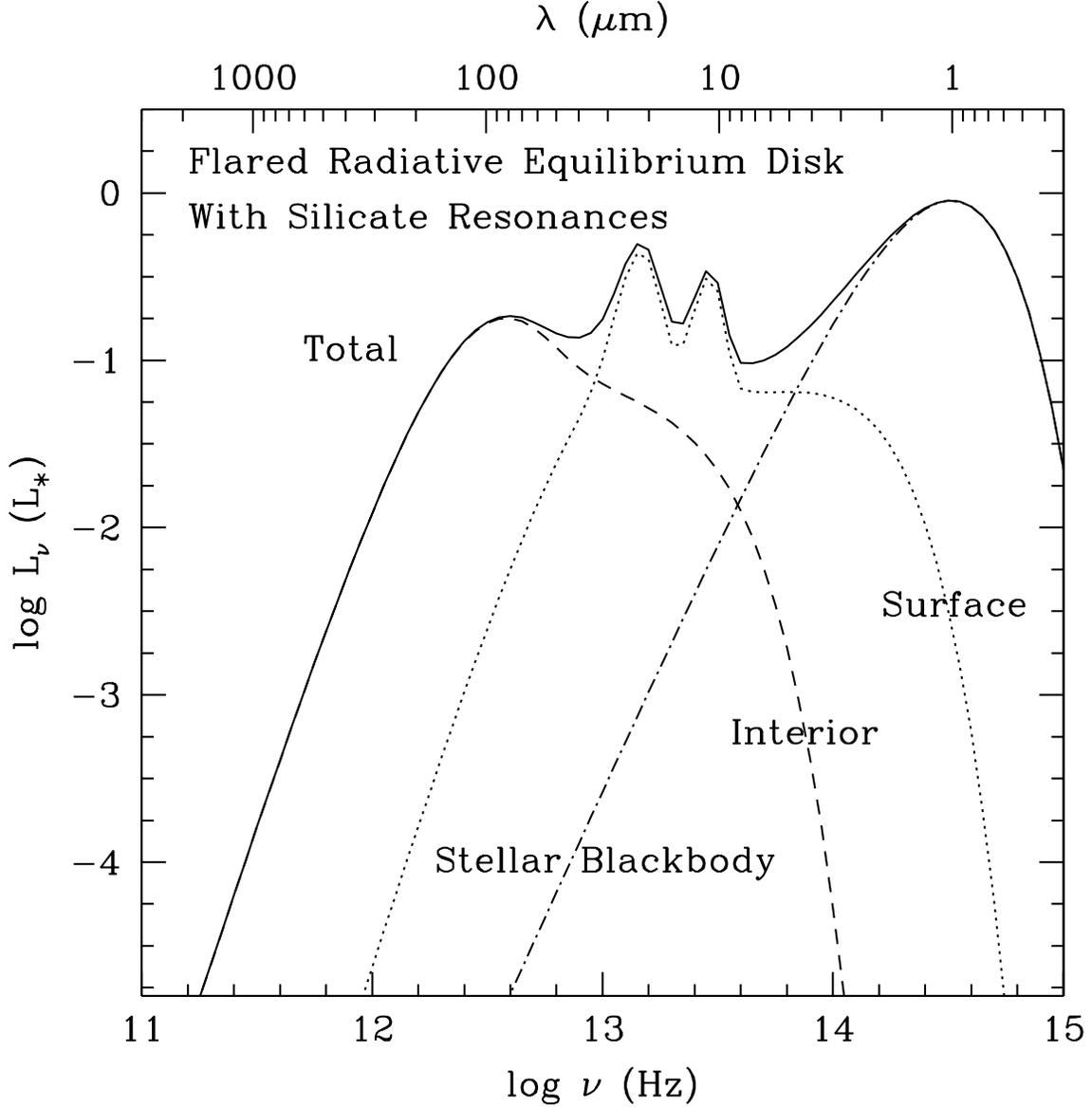}
\caption{SED for the hydrostatic, radiative equilibrium disk using a grain
emissivity profile
motivated by data from \protect{\markcite{m90}} Mathis (1990). For wavelengths
shorter
than $0.3\mum$, our assumed emissivity is unity; longward
of $0.3\mum$, it obeys a (single) power law relation $\varepsilon _{\lambda} =
(0.3\mum/\lambda)^{1.4}$,
on which are superposed two Gaussians centered on 10 and $20\mum$, having
amplitudes
that are 3 times their local continuum emissivity and FWHM equal to 3 and
$9\mum$, respectively. \label{specd}}
\end{figure}

Emission features are
a natural consequence of the superheated dust layer for disks viewed nearly
face-on. Line to continuum ratios
should be of the same order
as the percentage increase in dust emissivity associated with the
spectral feature. To illustrate the effects of spectral resonances, we
have employed an emissivity profile motivated by the data for ``outer-cloud
dust''
as given by \markcite{m90} Mathis (1990).
The resulting (face-on) SED displayed in Figure \ref{specd} exhibits
emission features at 10 and $20\mum$.

Absorption features may be associated with disks viewed nearly
edge-on. Support for this hypothesis comes from two observational
findings. These features appear most prominently among the
``extreme'' flat spectrum sources for which $L_{\nu}$ at infrared
wavelengths matches or exceeds $L_{\nu}$ in the visible. There is a
positive correlation between the strengths of these features and the
fractional linear polarization at optical wavelengths. Both
findings are plausibly the consequence of the extinction of optical
radiation in a flared disk viewed at large inclination angle.

\placefigure{specd}

\subsubsection{Temporal Behavior}

Temporal variations would provide another diagnostic
of circumstellar disks. Of relevance here is the claim by
\markcite{mb97} Moriarty-Schieven \& Butner (1997) that the submillimeter
and millimeter fluxes from the T Tauri binary GG Tau increased by factors of
order 2 between 1992 and 1994. The cause of this ``radio-wave flare'' has
not been identified. A plausible hypothesis is that it resulted from
enhanced disk heating associated with a burst in luminosity originating
near one or both components of the central binary. This leads us to consider
relevant timescales for the radiative and hydrostatic response of the disk.

Seven different timescales come into play. They are: (1) the timescale over
which superheated dust grains
in the surface layer equilibrate with the ambient stellar radiation field,

\begin{equation}
t_{ds}\sim \frac{r\rho_d kT_{ds}a^2}{\mu_d R_*^2 \sigma T_*^4}\sim 0.02
a_\au^{8/5} \s \, ,
\label{tds}
\end{equation}

\noindent where $\mu_d \approx 10\mu_g$ is the mean molecular weight per degree
of freedom in
a dust grain; (2) the light-travel
timescale from star to disk;

\begin{equation}
t_{lt}\sim \frac{a}{c}\sim 5 \times 10^2 a_\au \s\; ;
\label{tlt}
\end{equation}

\noindent (3) the photon diffusion timescale through the disk,\footnote{This
timescale is only relevant
where the disk interior is opaque to its thermal radiation.}

\begin{equation}
t_{di\!f\!f}\sim \frac{\varepsilon_i\kappa_V\Sigma h}{c}\sim \frac{2 \times
10^5}{a_\au^{9/14}}\s\, ;
\label{tpd}
\end{equation}

\noindent (4) the thermal timescale for the dust if it is decoupled from the
gas,\footnote{In evaluating the timescale at the outermost radii, we have set
$T_i = 21 K$.}

\begin{equation}
t_{di}\sim \frac{\Sigma_d k}{\mu_d\sigma
T_i^3}\left(1+\frac{1}{\varepsilon_i\kappa_V\Sigma}\right)
\sim \frac{2\times 10^5}{a_\au^{3/14}} \s \times \left\{ \begin{array}{ll}
(1 + 10^{-4}a_\au^{27/14})  & \mbox{for $a_\au \lesssim 84$} \\
(1 +\left(\frac{a_\au}{84}\right)^{3/2}) & \mbox{for $84 \lesssim a_\au
\lesssim 270$}
	\end{array} \right. \, ,
\label{tdi}
\end{equation}

\noindent where $\Sigma_d\approx 10^{-2}\Sigma$ is the surface density of dust;
(5) the thermal timescale for the
gas,\addtocounter{footnote}{-1}$\mbox{\footnote{}}$

\begin{equation}
t_{gi}\sim \frac{\Sigma k}{\mu_g\sigma T_i^3}
\left(1+\frac{1}{\varepsilon_i\kappa_V\Sigma}\right)
\sim \frac{2\times 10^8}{a_\au^{3/14}} \s \times \left\{  \begin{array}{ll}
(1 + 10^{-4}a_\au^{27/14})  & \mbox{for $a_\au \lesssim 84$} \\
(1 +\left(\frac{a_\au}{84}\right)^{3/2}) & \mbox{for $84 \lesssim a_\au
\lesssim 270$}
	\end{array} \right. \, ;
\label{tgi}
\end{equation}

\noindent (6) the timescale for the dust temperature to relax to the
gas temperature,

\begin{equation}
t_{relax}\sim \frac{r\rho_d}{\mu_d n_g v_g}\sim \frac{\mu_g r\rho_d}
{\mu_d\Sigma\Omega}\sim 10^{-2} a_\au^3\s\, ;
\label{tdg}
\end{equation}

\noindent (7) the dynamical timescale over which the disk adjusts
to departures from hydrostatic equilibrium,

\begin{equation}
t_{dyn}\sim 1.4a_\au^{3/2}\yr\, .
\label{tdyn}
\end{equation}

How rapidly might the SED vary in response to changes in the luminosity of
the central star? Since $t_{ds}\ll t_{lt}$, contributions
from the surface layer are limited by $t_{lt}$. Those from
the interior are limited by $t_{gi}$, since $t_{relax}\lesssim t_{di}$ and
$t_{di\!f\!f}\ll t_{gi}$.\footnote{The former inequality is only
marginally satisfied at $a_o$.} To relate these response times to timescales
for variation at a fixed wavelength, $\lambda$, consult Figure \ref{fluxfig}.

\subsection{Unresolved Issues}

Our investigation leaves many unresolved issues.

Is a disk in radiative and hydrostatic equilibrium dynamically stable?

How does the SED depend upon disk inclination?

How much of the
thermally-emitted spectrum is covered by molecular lines?  Do the lines appear
in absorption or emission? Which molecules are they associated with?

How high does the layer of superthermal dust grains sit above the
disk midplane? What is the gas temperature in this layer?  Is gravitational
settling of dust grains seriously impeded by turbulent mixing or by vertical
circulation? Or
does radiation pressure from the central star dominate all the forces, so that
the superthermal grains are driven into denser environments and the disk
geometry
flattens?

What is the albedo of the circumstellar dust? How does scattered stellar
radiation compare to thermal emission in the near IR?

Does active accretion significantly affect the SEDs, either by
preventing the settling of dust grains or by thickening the inner
regions of the disk?

\acknowledgments
We thank Anneila Sargent for providing the data
on GM Aur and other T Tauri stars, Shri Kulkarni for suggesting
that temporal variability of disk emission can provide a further
means of probing disk characteristics, Dave Koerner and Karl Stapelfeldt
for informative discussions regarding GM Aur, and Steve Beckwith for
a thorough and thoughtful referee's report.
Financial support for this research was
provided by NSF Grant 94-14232.
E. C. acknowledges support from an NSF Graduate Fellowship.


\begin{references}
\reference{als87} Adams, F. C., Lada, C. J., \& Shu, F. H. 1987, \apj, 312, 788
\reference{ars89} Adams, F. C., Ruden, S. P., \& Shu, F. H. 1989, \apj, 347,
959
\reference{betal90} Beckwith, S. V. W., Sargent, A. I., Chini, R. S., \&
Gusten, R. 1990, \aj, 99, 924
\reference{cetal94} Calvet, N., Hartmann, L., Kenyon, S. J., \& Whitney, B. A.
1994, \apj, 434, 330
\reference{cw85} Cohen, M., \& Witteborn, F. C. 1985, \apj, 294, 345
\reference{d95} Draine, B. T. 1995, in The Physics of the Interstellar Medium
and Intergalactic Medium, ed. A. Ferrara, C. F. McKee, C. Heiles, \& P. R.
Shapiro. A. S. P. Conf. Series, 80, 133
\reference{dl84} Draine, B. T., \& Lee, H. M. 1984, \apj, 285, 89
\reference{kh87} Kenyon, S. J., \& Hartmann, L. 1987, \apj, 323, 714
\reference{h96} Hunter, T. 1996, PhD Thesis
\reference{k97} Koerner, D. W. 1997, Origins of Life and Evolution of the
Biosphere, 27, 157
\reference{k70} Kusaka, T., Nakano, T., \& Hayashi, C. 1970, Prog. Theo. Phys.,
44, 1580
\reference{letal94} Lay, O. P., Carlstrom, J. E., Hills, R. E., \& Phillips, T.
G. 1994, \apj, 434, L75
\reference{lp74} Lynden-Bell, D., \& Pringle, J. E. 1974, \mnras, 168, 603
\reference{m90} Mathis, J. M. 1990, \araa, 28, 37
\reference{mo96} McCaughrean, M. J., \& O'Dell C. R. 1996, \aj, 111, 1977
\reference{m68} Mendoza, E. E. V. 1968, \apj, 151, 977
\reference{mb97} Moriarty-Schieven, G. H., \& Butner, H. M. 1997, \apj, 474,
768
\reference{metal96} Mundy, L. G., Looney, L. W., Erickson, W., Grossman, A.,
Welch, W. J., Forster, J. R., Wright, M. C. H., Plambeck, R. L., Lugten, J., \&
Thornton, D. D. 1996, \apj, 464, L169
\reference{n93} Natta, A. 1993, \apj, 412, 761
\reference{osa92} Ostriker, E. C., Shu, F. H., \& Adams, F. C. 1992, \apj, 399,
192
\reference{r85} Rucinski, S. M. 1985, \aj, 90, 2321
\reference{rp91} Ruden, S. P., \& Pollack, J. B. 1991, \apj, 375, 740
\reference{retal84} Rydgren, A. E., Schmelz, J. T., Zak, D. S., \& Vrba, F. J.
1984, Broad Band Spectral Energy Distributions of T Tauri Stars in the
Taurus-Auriga Region, Pub. US Naval Obs., 2nd Ser. Vol. 25, 1
\reference{rz87} Rydgren, A. E., \& Zak, D. S. 1987, \pasp, 99, 141
\reference{setal89} Strom, K. M., Strom, S. E., Edwards, S., Cabrit, S., \&
Skrutskie, M. F. 1989, \aj, 97, 1451
\reference{sal87} Shu, F. H., Adams, F. C., \& Lizano, S. 1987, \araa, 25, 23
\reference{setal90} Shu, F. H., Tremaine, S., Adams, F. C., \& Ruden, S. P.
1990, \apj, 358, 495
\reference{w77} Weidenschilling, S. J. 1977, \apss, 51, 153
\end{references}
\end{document}